# Hydrogen Storage Potential of Chromium-Functionalized Graphene: A First-Principles Investigation

Pratyasha Tripathy[1], Hetvi Jadav[2], Himanshu Pandey[2*]

[1]*Department of Physical Sciences, Indian Institute of Science Education and Research, Berhampur, Odisha 760010, India*

[2]*Condensed Matter & Low-Dimensional Systems Laboratory, Department of Physics, Sardar Vallabhbhai National Institute of Technology, Surat 395007, India*

## Abstract

Sorbent materials, such as graphene-based systems coated with Cr, are being investigated as potential hydrogen storage materials. Graphene, a 2D material with a high surface-to-volume ratio, has been employed. A comparison is conducted between graphene systems with single vacancy defects and those without defects, using Cr adsorption. To verify the effectiveness of hydrogen storage, *ab initio* calculations are carried out both with and without Van der Waals interactions. The system's binding energy is calculated to assess efficiency. According to the Department of Energy in the United States, the ideal range for binding energy for reversible hydrogen storage is between 0.2 and 0.6 eV. To anticipate the stability of the efficient materials at room temperature, this work exploits the molecular dynamics computations to depict their thermal stability spectrum.

## 1 Introduction

There has been a significant urge to explore alternative energy sources to cater to the requirements of the growing population, which aids in high urban expansion, with primary energy sources being fossil fuels [1,2]. These primary resources are parents to greenhouse gases, primarily carbon dioxide, which have led to severe environmental implications, including global warming, rising sea levels, and biodiversity loss [2,3]. The sudden and several incidents of extreme weather conditions, such as melting of glaciers, storms, droughts, heavy rainfall in a short duration, and floods, are also directly linked to greenhouse gas emissions. Furthermore, other by-products like sulphur dioxide, nitrogen oxides, and particulate matter contribute to air pollution, causing major health disorders, including respiratory and cardiovascular diseases [2,4]. Additionally, 80% of the world's energy contribution is attributed to fossil fuels, raising concerns due to the uneven distribution of fossil fuel reserves, leading to economic and geopolitical scrutiny.

This scenario has catalysed the search for cleaner and more sustainable energy sources [5]. Current renewable sources include natural sources like wind and solar energy, with low or negligible greenhouse gas emissions and the least environmental impact. These sources can lower the global carbon footprint significantly [6]. Renewable energy systems provide a wide range of benefits, positioning them as a sustainable and environmentally responsible alternative to fossil fuels. By harnessing natural resources such as sunlight, wind, water, and geothermal heat, these systems generate clean energy with little to no greenhouse gas emissions. This significantly reduces our reliance on carbon-intensive energy sources, thereby helping to mitigate climate change and lessen its detrimental impacts, such as rising global temperatures, extreme weather events, and sea level rise. In addition to their environmental advantages, renewable energy systems also promote energy security, job opportunities, and support economic growth. The sector also drives economic growth by creating millions of jobs worldwide. Unlike fossil fuels, which are finite and often subject to volatile global markets, renewable resources are abundant and locally available, offering long-term energy stability and resilience. Furthermore, renewable energy systems have lower operational costs and diverse applications in electricity generation, heating, and transportation, making them a versatile and cost-effective solution for the future [2,7].

Hydrogen has reaped noteworthy attention as a capable candidate for green energy solutions [8-11]. Hydrogen is abundant in nature and is considered an efficient renewable energy source. The combustion of hydrogen leads to the formation of water as a by-product instead of carbonaceous compounds, which are the leading contributors to greenhouse gas emissions [12]. However, transitioning to a hydrogen-based economy requires significant technological advancements, including production, transportation, storage, and utilization of hydrogen. Among these, the storage of hydrogen is still one of the most challenging issues due to its low volumetric energy density, requiring innovative solutions to make large-scale hydrogen energy implementation viable [13].

Hydrogen technology has been highly utilized in fuel cell production with rising energy demands. Fuel cells use hydrogen as a source, undergoing electrochemical processes involving oxygen to produce energy [14-17]. Different types of fuel cells have proven to be more efficient than traditional energy-producing sources. Solid oxide fuel cells [2,18] and molten carbonate fuel cells [ 2,19] exhibit high adaptability and have great potential for combined heat and power generation. Similarly, microbial fuel cells [20] have the potential to convert direct waste into energy. In contrast, fuel cells involving polymer membranes or proton-conducting ceramic materials [21] can function at lower temperatures, making them highly suitable for portable and automotive applications. Hydrogen technology also finds utility in gas turbines and hydrogen internal combustion engines [22], which use ignition engines such as spark-ignition [23] and compression-ignition [24] engines, where hydrogen combustion is the primary principle for energy harnessing. Although hydrogen is abundant in nature and an efficient renewable energy source, current production methods can be categorized into four types: thermochemical, electrochemical, biological, and photocatalytic production [25]. The thermochemical [26] and electrochemical methods [27] for hydrogen production are well-established, while the latter two are still in the developmental stages due to low yields and material constraints. A significant drawback of well-established thermochemical methods is that they often lead to greenhouse gas emissions, which contradicts the goal of clean energy. Therefore, developing efficient and sustainable hydrogen production methods is a critical area of ongoing research. However, this study primarily focuses on hydrogen storage, a crucial bottleneck in hydrogen energy applications.

Hydrogen storage technology can be categorized into two groups: physical and material-based storage [28]. Physical storage methods include compressed gas storage, cryo-compressed gas storage, and liquid hydrogen-based storage [28]. Although compressed gas storage is suitable for low energy demands and storing high-purity hydrogen, it requires cumbersome high-pressure tanks with low energy density [29]. Similarly, although cryogenic liquid hydrogen storage provides high energy density, it comes with notable challenges. These include the need to maintain extremely low temperatures, high energy consumption during the liquefaction process, and losses caused by boil-off and evaporation [30-31]. These challenges necessitate the development of advanced materials for hydrogen storage, leading to extensive research into material-based storage methods. Material-based storage methods have gained significant research interest due to their higher efficiency and potential for practical applications. This research work focuses on studying adsorbent materials for efficient hydrogen storage. Recent research on transition metal-based 2D materials [32-34] and carbon nanotubes [35] has shown promising potential for hydrogen storage [12]. Here, transition metals like chromium (Cr) decorated graphene systems are explored, considering defect-free graphene and graphene with single vacancy defects [36]. Pristine graphene alone does not exhibit a preferred binding energy for hydrogen adsorption. However, graphene's high surface-to-mass ratio and low weight make it a promising candidate for storage applications [33]. Decorating graphene with transition metals enhances its hydrogen storage capacity by modifying its electronic structure and increasing adsorption sites.

To investigate these properties, first-principles calculations using Density Functional Theory (DFT) are carried out [37, 38]. The binding energy for hydrogen adsorption and the stable position for transition metals are computed to determine the most effective configurations. Cr-decorated graphene sheets are analyzed for defect-free and single vacancy defect systems, allowing for a comparative study of the impact of defects on hydrogen storage performance [12,36]. Furthermore, the effect of Van der Waals forces on the system is investigated to get a more realistic assessment of hydrogen adsorption behaviour [39]. Hydrogen storage efficiency is evaluated by calculating the system's binding energy, a crucial parameter for reversible hydrogen storage applications. According to the United States Department of Energy (DOE) [28], the preferred binding energy range for effective hydrogen storage is between 0.2 and 0.6 eV [12, 33, 40]. DFT calculations are performed at absolute zero temperature (0 K), which does not fully account for the thermal effects present under real-world conditions. To predict the stability of efficient materials at room temperature, molecular dynamics (MD) simulations are carried out in a canonical ensemble at 300 K. MD simulations [41-43] provide insights into the material's response to thermal fluctuations, and the thermal stability spectrum is analysed to determine whether Cr-decorated graphene remains effective under ambient conditions.

This study aims to provide valuable insights into the feasibility of Cr-decorated graphene as a potential material for hydrogen storage. The findings contribute to the ongoing efforts to develop advanced hydrogen storage solutions, supporting the broader goal of transitioning toward a sustainable hydrogen economy. Future research directions include exploring alternative transition metals, optimizing synthesis techniques, and further integrating experimental validation with computational studies to enhance hydrogen storage efficiency.

## 2 Computational Details

A plane-wave-based DFT implemented in the Quantum Espresso (QE) [44] package was employed for all calculations. GGA (generalized gradient approximation) [45] based exchange-correlation functional, as described by PBE, was utilized. An ultra-soft pseudo-potential is used for all computations. In a 3x3 supercell, the graphene structure is optimized. For the Cr-decorated graphene system, the height optimization calculation is done. The exact process is followed for hydrogen molecule adsorption. The van der Waals correction factor (vdw corr) [39] is included in the computation by involving the dispersion force correction factor as in the DFT-D3 method Grimme gave. All the structures are visualised using Xcrysden [46]. Each calculation has an energy convergence threshold of $10^{-8}$ Ry with a grid of 4x4x1 *kpoints* for a defect-free graphene system. Further, the kinetic energy cutoff for wavefunction (*ecutwfc*) and charge density (*ecutrho*) is 130 Ry and 1300 Ry, respectively. Further, the distance between the graphene layers is kept fixed at 10 Å to prevent any interaction between the layers. Considering a single vacancy defect for the graphene system, for each calculation, the *ecutwfc* and *ecutrho* parameters were set to 90 Ry and 900 Ry, respectively. The de-gauss values for all the systems are considered to be 0.01. The dynamic stability of the systems was calculated using Car-Parinello molecular dynamics in the canonical system (NVT) in QE. The system for molecular dynamics calculations is considered to be in 3x3 supercells, having 20 atoms in a Cr-adsorbed single vacancy defect Graphene system with $H_2$ adsorbed on it. The nose thermostat is run at 300 K to study the system's stability at room temperature. The computations are performed for 55,000 steps at a time step of 4 au.

## 3 Results and Discussion

### 3.1 Defect-free graphene (DFG)

The defect-free graphene monolayer is created by taking a 3x3 supercell of 18 atoms. The geometry optimization computation is performed to determine the bond length and the optimum lattice constant. The monolayer can be visualized from Fig. 1. A hexagonal unit cell with the lattice parameter 2.468 Å is taken. The bond length is measured to be 1.426 Å, close to the previous experimental observation of 1.424 Å [12,47]. In this graphene monolayer, Cr metal can be adsorbed at one of the three positions available, *i.e.*, top (T), bridge (B), or hollow (H) positions as shown in Fig. 1. For the top carbon site, binding energy values are -1.738 eV and -0.441 eV, respectively, with and without the van der Waals force. When we tried to adsorb a Cr atom on the bridge site, it moved to the hollow site to achieve stability. The most stable position was found to be at the hollow position, as in Fig. 2. The height for adsorption of Cr metal is calculated for both cases with and without the van der Waals correction factor. The binding energy for the Cr metal is computed using $(E_b)_{Cr} = E_{DFG-Cr} - E_{DFG} - E_{Cr}$, where $(E_b)_{Cr}$ is the binding energy of Cr, $E_{DFG-Cr}$ is the energy of the Cr adsorbed on a defect-free Graphene system, $E_{DFG}$ is the energy of defect-free graphene, and $E_{Cr}$ is the energy of Cr in its crystal structure.

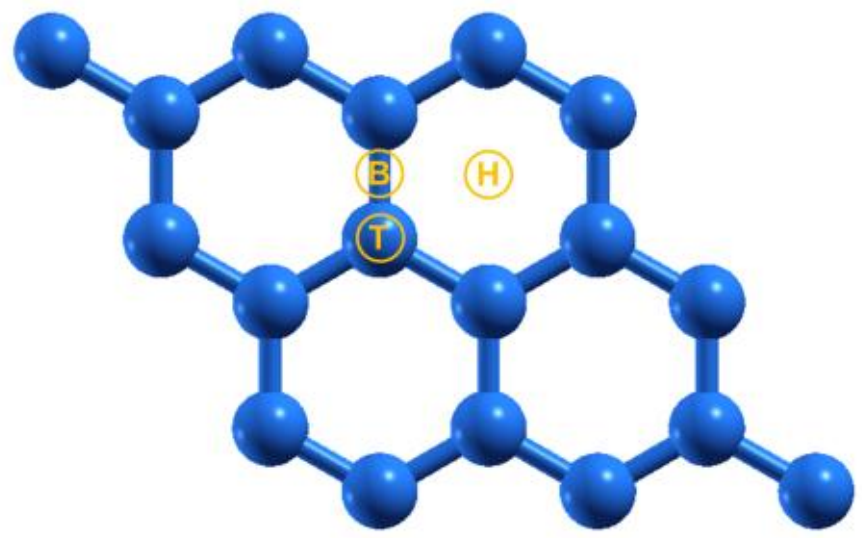

**Figure 1** Defect Free Graphene (DFG) Monolayer, Three possible adsorption sites; H = hollow site, T = top of carbon site, B = bridge site.

The height for Cr adsorbed when the van der Waals force is included is 1.516 Å. The results are summerized in Table 1, but when the system is relaxed without the Van der Waals force, presuming the predominance of the Kubas interaction, the height at which Cr gets adsorbed is 1.514 Å. The binding energy for Cr to be adsorbed on the graphene monolayer is -3.17 eV when van der Waals force is predominant and -1.89 eV when Kubas interaction is predominant. For both cases, the binding energy suggests the stable nature of the Cr-decorated graphene monolayer, which can be found in Table 2. Thus, this system can be further studied for hydrogen adsorption purposes. When a hydrogen molecule is absorbed on the Cr-adsorbed graphene, we must determine whether this is feasible. For that, we have calculated the binding energy for the whole system. The binding energy of the $H_2$ molecule is calculated using Eq. (1).

$$(E_b)_{H_2} = E_{DFG-Cr-H_2} - E_{DFG-Cr} - E_{H_2} \quad (1)$$

where $(E_b)_{H_2}$ is the binding energy of $H_2$, $E_{DFG-Cr-H_2}$ is the energy of the system after hydrogen adsorption, $E_{DFG-Cr}$ is the energy of the graphene monolayer after Cr adsorption, and $E_{H_2}$ is the energy of $H_2$ optimized in its experimentally used crystal structure.

The height of $H_2$ after adsorption is calculated. The height for $H_2$ adsorption when the Van der Waals force is included is 1.697 Å. But when the system is relaxed without Van der Waal force, presuming the predominance of Kubas interaction [33], the height at which Cr gets adsorbed is 1.699 Å. The results are listed in Table 3. The binding energy for DFG-Cr-$H_2$ is further calculated. It is found to be -1.03 eV when the Van der Waals force is predominant and -0.945 eV when the Kubas interaction is predominant. The results are listed in Table 4. For the DFG-Cr system, the binding energy of $H_2$ is a bit higher than the preferred range suggested by the DOE for reversible hydrogen storage. Still, it has the potential to be used to transport $H_2$ over a significant distance. Thus, Cr-decorated single vacancy defect graphene is further explored for hydrogen storage [48].

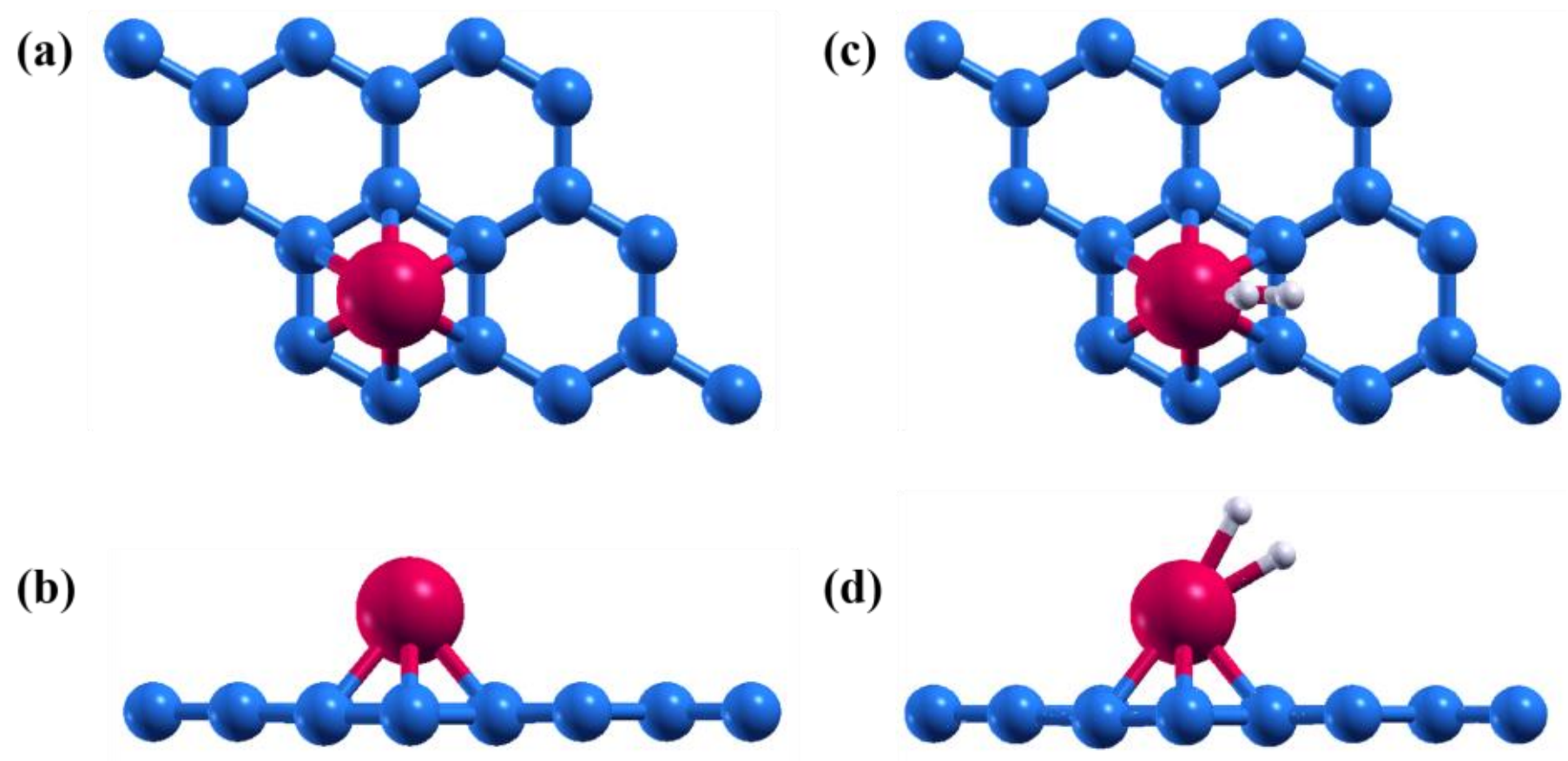

**Figure 2** (a) Top view and (b) Side view of DFG-Cr system at hollow site (site-H, (c) Top view and (d) Side view of DFG-Cr-$H_2$ system

## 3.2 Single Vacancy Graphene (SVG)

The Single Vacancy graphene (SVG) monolayer is considered with a 3x3 supercell consisting of 17 atoms. The monolayer can be visualized from Fig. 3. The system is supposed to be in a subsequent layer. The bond length is measured to be 1.426 Å, per the experimental data (1.424 Å) [12,47]. To study the behaviour of the Cr atom on SVG, we adsorbed Cr on the monolayer. Cr can be adsorbed on seven available positions on the SVG monolayer, *i.e.*, T = top of carbon site, B = top of bridge site, H = hollow site, V = vacancy site and VT = top of carbon near vacancy, VB = bridge site near vacancy, and VH = hollow positions near the vacancy, which are shown in Figure 3. The most stable position for the Cr atom to be adsorbed is at the vacancy position (V-site). These results were obtained by running optimization calculations on the system with and without including the Van der Waals' correction factor. The height at which Cr is adsorbed when the Van der Waals force is predominant is 1.24 Å. In the case when Kubas' interaction is predominant, the height is found to be 1.25 Å. The results are listed in Table 1. The binding energy of the system can be calculated using Eq. (2) given below.

$$(\mathrm{E_b})_{\mathrm{Cr}} = \mathrm{E_{SVG-Cr}} - \mathrm{E_{SVG}} - \mathrm{E_{Cr}} \tag{2}$$

Where $(\mathrm{E_b})_{\mathrm{Cr}}$ is the binding energy of Cr on the SVG system, $E_{SVG-Cr}$ is the system's energy after Cr adsorption, $E_{SVG}$ is the energy of the single vacancy graphene monolayer, and $E_{Cr}$ is the energy of the Cr atom. The binding energy is calculated to be -9.57 eV when van der Waals force is predominant and -8.42 eV when Kubas interaction is predominant. The binding energies of the system predict its stable nature and are summerized in Table 2. The system is further studied for hydrogen storage by studying hydrogen adsorption.

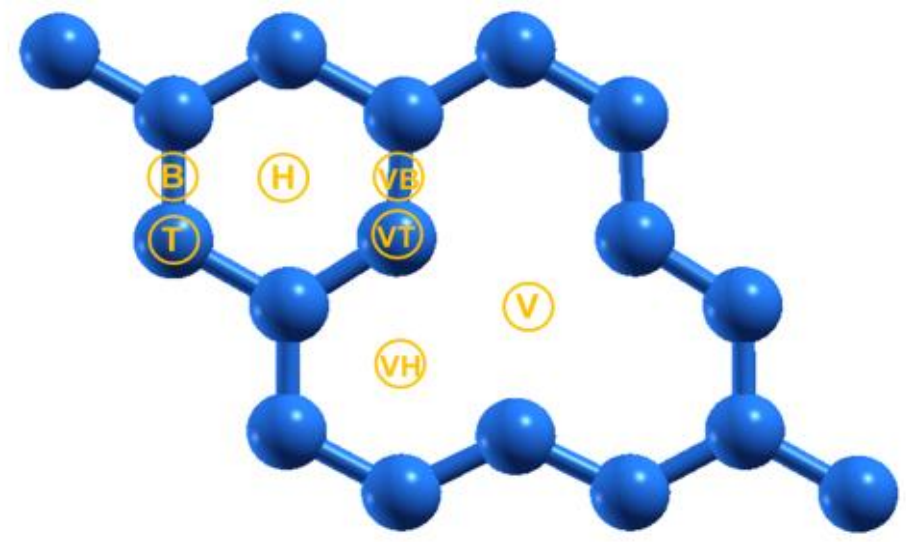

**Figure 3** Single Vacancy Graphene with possible six adsorption sites (i) V = Vacancy (ii) VH = Vacancy Hollow (iii) VT = Vacancy Top (iv) VB = Vacancy Bridge (v) H = hollow site (vi) T = top site (vii) B = bridge site

The hydrogen molecule is adsorbed on the surface of the SVG-Cr system. The adsorption height is determined using geometry optimization calculations, for the system including Van der Waals force, the adsorption height for $H_2$ is found to be 1.794 Å and 1.807 Å for that with Kubas interaction. Further, the binding energy for the systems is also calculated. For the former case, it is -0.44 eV and -0.38 eV for the latter case.

**TABLE-1:** Height of Cr after Adsorption on Graphene and height of $H_2$ Adsorption on Graphene-Cr-$H_2$ system with and without Van der Waals correction

| TRANSITION METAL | Height with Van der Waals correction (Å) | Height without Van der Waals correction (Å) |
|---|---|---|
| DFG-Cr | 1.516 | 1.514 |
| SVG-Cr | 1.240 | 1.250 |
| DFG-Cr-$H_2$ | 1.697 | 1.699 |
| SVG-Cr-$H_2$ | 1.794 | 1.807 |

**TABLE-2-:** Binding energy of the Cr adsorbed Graphene system and Graphene-Cr-$H_2$ system

| TRANSITION METAL | Binding Energy with van der Waals correction (eV) | Binding Energy without Van der Waals correction (eV) |
|---|---|---|
| DFG-Cr | -3.17 | -1.89 |
| SVG-Cr | -9.57 | -8.42 |
| DFG-Cr-$H_2$ | -1.03 | -0.945 |
| SVG-Cr-$H_2$ | -0.44 | -0.38 |

### 3.3 Desorption Temperature Calculation

To calculate the system's applicability and stability, the van't Hoff equation [49,50] is applied, which gives insight into dehydrogenation thermodynamics, especially the desorption temperature. The equation is as follows [49,51,52]:

$$T_d = \Delta H / \Delta S$$

Where $T_d$ represents the desorption temperature, $\Delta H$ and $\Delta S$ are the hydrogen desorption enthalpy and entropy. While $\Delta S$ for hydrogen desorption has been predetermined from previous works and is reported to be -130.7 J/K [51,53,54], ΔH can be calculated using the following equation using Hess's law [53]:

$$\Delta \mathrm{H} = \Sigma_{\mathrm{total}}\, \mathrm{E}_{\mathrm{products}} - \Sigma_{\mathrm{total}}\, \mathrm{E}_{\mathrm{reactants}}$$

The enthalpies for hydrogen adsorbed defect-free and single vacancy graphene are determined to be -1.03 eV and -0.44 eV, respectively, and the desorption temperatures are 764 K and 325 K, respectively.

### 3.4 Molecular Dynamics on SVG-Cr-$H_2$ system

The estimated binding energy of the Cr metal-adsorbed systems is within the energy range specified by the DOE standards for reversible hydrogen storage. Although these systems have been studied using DFT and have shown promising results, it is crucial to further investigate their stability at room temperature to predict their experimental feasibility, as DFT calculations typically assume the systems are at 0 K.

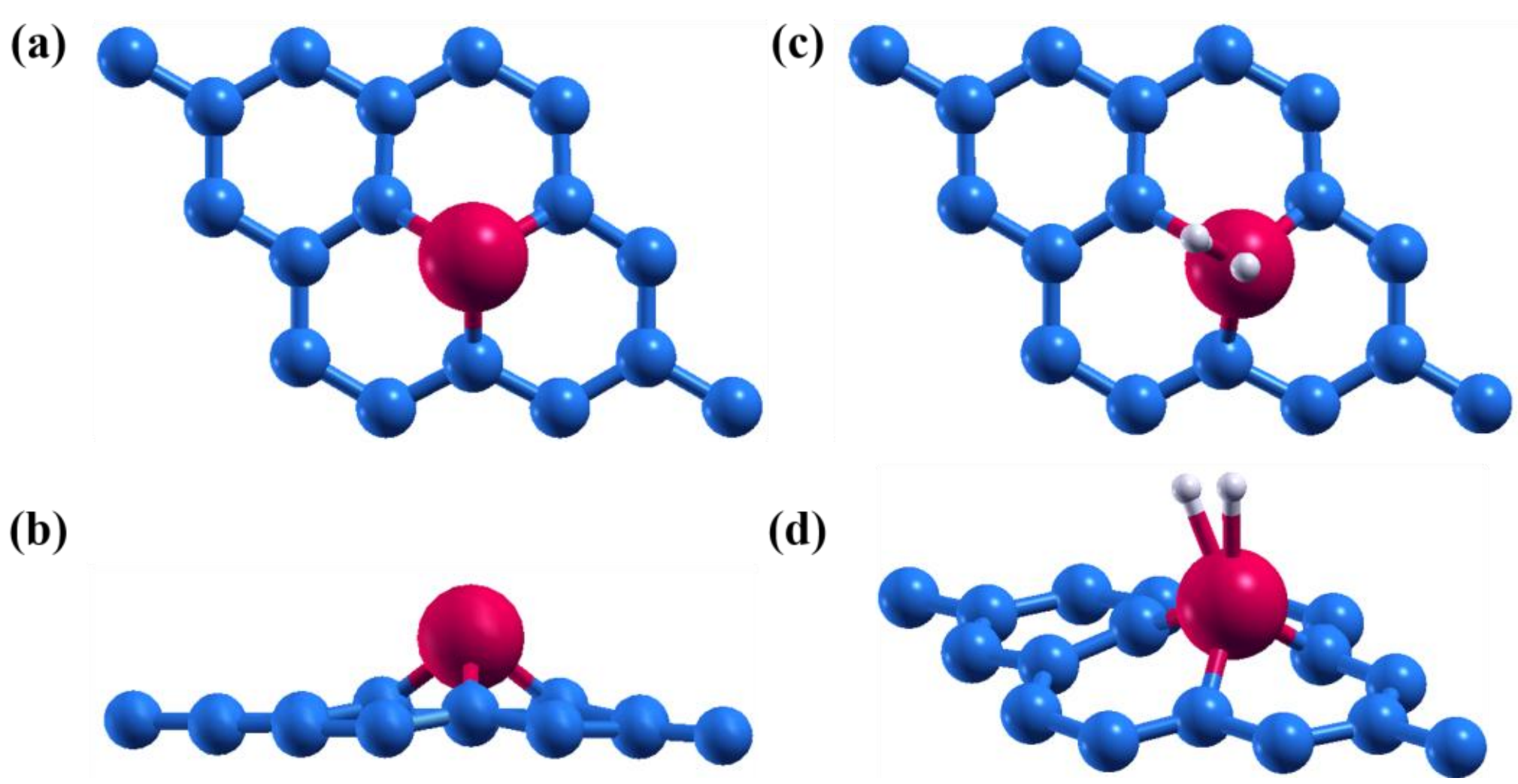

**Figure 4** (a) Top view and (b) Side view of SVGr-Cr system at hollow site (Site-H), (c) Top view and (d) Side view of SVG-Cr-$H_2$ system.

To address this, MD simulations were carried out to investigate the thermodynamic stability of these systems. While DFT operates under the assumption of an idealized environment at 0 K, predicting the experimental viability of the material requires confirmation of its stability at 300 K or elevated temperatures.

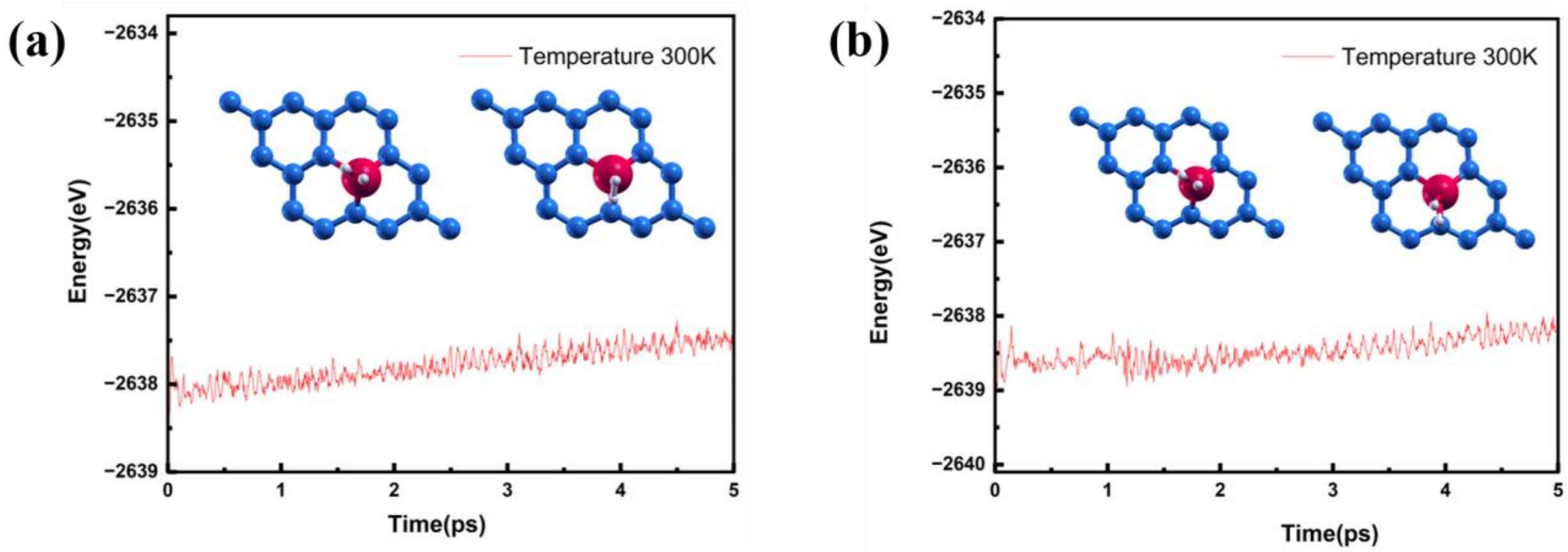

**Figure 5** Molecular Dynamics for SVG-Cr-$H_2$ (a) with and (b) without Van der Waals interaction

Car-Parrinello Molecular Dynamics (CPMD) calculations were conducted in this study, considering a canonical system with an effective mass of 300 atomic units (a.u.). The energy variation of the system was visualized using the thermostability spectrum over a period of 5 picoseconds (ps), with measurements taken at intervals of 4 a.u. Figures 5(a) and 5(b) illustrate the initial structure of both systems along with the structure obtained after running the MD calculations. The results show almost no structural change, indicating the stable nature of the system. This stability at room temperature suggests that the Cr metal-adsorbed system is experimentally viable and can maintain its integrity under realistic conditions, making it a probable candidate for hydrogen storage applications. The findings from the MD simulations complement the DFT results, reinforcing the potential of the Cr metal adsorbed system for practical hydrogen storage and transportation. The observed stability and favourable binding energy characteristics highlight the system's suitability for advancing hydrogen storage technologies and addressing the challenges associated with safe and efficient hydrogen transportation.

## 4 Conclusion

Throughout this research, we have determined that the chromium metal-adsorbed single vacancy system exhibits considerable potential for reversible hydrogen storage applications. The defect-free system, however, presents a higher binding energy than the optimal range. Notably, the binding energy for hydrogen storage exceeds the parameters set by the DOE, indicating that the defect-free Cr-adsorbed graphene system (high desorption temperatures of 764 K) can be employed for hydrogen storage with assured safety during transportation over long distances. This finding underscores the potential of Cr-adsorbed defect-free graphene systems for secure, long-distance hydrogen transport without the risk of leakage. Also, the Cr-adsorbed graphene sheet with a single vacancy is found suitable for routine hydrogen storage applications as its binding energy is 0.44 eV, within the DOE standards (low desorption temperatures of 325 K). Additionally, molecular dynamics simulations reveal that the system maintains its stability at room temperature.

The calculations confirm that the structural integrity and performance of the Cr-adsorbed graphene system are not compromised under ambient conditions, thereby reinforcing its suitability for practical hydrogen storage and transportation applications. The observed stability and favourable binding energy characteristics suggest that this system could significantly advance hydrogen storage technologies and address the challenges associated with safe and efficient hydrogen transportation.

## Acknowledgment

This research is supported by the research grant from the Institute Seed Research Grant 2021–22/DOP/04.